\documentclass[reprint,aps,prm,twocolumn,superscriptaddress,floatfix]{revtex4-1}
\usepackage{amsmath}
\usepackage{float}
\usepackage{placeins}
\usepackage[table]{xcolor}
\usepackage[hidelinks]{hyperref}
\usepackage{epsfig}
\usepackage{graphics}
\usepackage{braket}

\graphicspath{figure-factory/}
\begin{document}
\title{Modeling lithium-ion solid-state electrolytes with a pinball model}
\author{Leonid Kahle}
\author{Aris Marcolongo}
\author{Nicola Marzari}
\affiliation{Theory and Simulation of Materials (THEOS), and National Centre for Computational Design and Discovery of Novel Materials (MARVEL), \'{E}cole Polytechnique F\'{e}d\'{e}rale de Lausanne, CH-1015 Lausanne, Switzerland}
\date{\today}
\definecolor{mygreen}{rgb}{0.0, 0.5, 0.2}

\begin{abstract}
We introduce a simple and efficient model to describe the potential energy surface of lithium diffusing in a solid-state ionic conductor. 
First, we assume that the Li atoms are fully ionized and we neglect the weak dependence of the electronic valence charge density on the instantaneous position of the Li ions.
Second, we freeze the atoms of the host lattice in their equilibrium positions; consequently, also the valence charge density is frozen. We thus obtain a
computational setup (the ``pinball model'') for which extremely inexpensive molecular dynamics simulation can be performed. 
To assess the accuracy of the model, we contrast it with full first-principles molecular dynamics simulations performed either with a free or frozen host lattice; in this latter case, the charge density still readjusts itself self-consistently to the actual positions of the diffusing Li ions.
We show that the pinball model is able to reproduce accurately the static and dynamic properties of the diffusing Li ions -- including forces, power spectra, and diffusion coefficients -- when compared to the self-consistent frozen-host lattice simulations.
The frozen-lattice approximation itself is often accurate enough, and certainly a good proxy in most materials. These observations unlock efficient ways to simulating the diffusion of lithium in the solid state, and provide additional physical insight into the respective roles of charge-density rearrangements or lattice vibrations in affecting lithium diffusion.
\end{abstract}

\maketitle

\section{Introduction}
Overcoming safety challenges and reaching performance targets in rechargeable Li-ion batteries will be key to the deployment of mobile applications such as electric vehicles~\cite{armand_building_2008}. 
The electrically insulating, ion-conducting electrolyte is a critical component in the quest to improve the power density, time stability and safety of batteries~\cite{quartarone_electrolytes_2011, balakrishnan_safety_2006}, and replacing the current state-of-the-art organic electrolytes with solid-state electrolytes (SSEs) offers an attractive alternative~\cite{kato_high-power_2016,manthiram_lithium_2017}. 
Despite the urgent need of new SSEs, only a small number of crystal structures with a sufficient ionic conductivity~\cite{bachman_inorganic_2016} have been discovered so far, and large regions of materials' space remain unexplored, highlighting the need and opportunity to find efficient ways to screen experimental or theoretical repositories of crystal structures for good ionic conductors. 
Accurate first-principles molecular dynamics (FPMD) simulations of diffusion properties are resource-limited to a few selected cases.
The first FPMD simulations of fast-ion or superionic conductors date to 1999~\cite{cavazzoni_superionic_1999} ($\mathrm{H_2O}$ and $\mathrm{NH_3}$) and 2006~\cite{wood_dynamical_2006} ($\mathrm{AgI}$), with few works specifically tackling Li-ion diffusion and Li-ion migration barriers with static~\cite{van_der_ven_first-principles_2001,du_ion_2007, holzwarth_computer_2011, lepley_structures_2013,du_structures_2014, lang_lithium_2015} and dynamic~\cite{  ong_phase_2012, mo_first_2012,xu_one-dimensional_2012, mo_insights_2014, meier_solid-state_2014, wang_design_2015, chu_insights_2016, zhu_li3y(ps4)2_2016, marcolongo_ionic_2017} first-principles methods.
More efforts employ classical force-fields to study diffusion phenomena in specific crystal families, such as the garnets~\cite{adams_ion_2012, xu_mechanisms_2012, kozinsky_effects_2016, klenk_finite-size_2016, burbano_sparse_2016} and the lithium superionic conductors (LISICONs)~\cite{adams_structural_2012, deng_structural_2015},
 but empirical methods often lack the generality to deal with large compositional variety.

In addition to direct dynamical simulations, several descriptors or design principles for ionic conductivity in solid-state materials have been suggested. 
As a first example, Wakamura and Aniya correlated optical phonon frequencies and the activation energy for diffusion in selected classes of materials~\cite{wakamura_roles_1997, wakamura_effects_1998}.
Another example is the importance of accessible volume for the diffusing species, which has been confirmed by experiments and simulations~\cite{kummer_-alumina_1972, adams_bond_2002}. This observation resulted in the bond-valence method~\cite{adams_determining_2000} which accounts, in a static single-particle picture, for volume and energy effects and has already been used in large-scale screening for ionic conductors~\cite{avdeev_screening_2012,xiao_candidate_2015}.
Wang et al.~\cite{wang_design_2015} could relate superionicity to the bcc-like topology of the underlying anionic sublattice, as was also discussed by Wood and Marzari~\cite{wood_dynamical_2006} for $\mathrm{AgI}$.
Work by Adelstein and Wood~\cite{adelstein_role_2016} showed how the mixed ionic-covalent nature of lithium bonds and frustration of the bonding during transition can explain superionic behavior observed in $\mathrm{Li_3InBr_6}$.
In addition to the search for descriptors, very recent work also highlights the importance of the collective nature of superionic diffusion~\cite{kozinsky_effects_2016, he_origin_2017, marcolongo_ionic_2017}.
An emerging trend is to tackle the descriptor search with machine-learning~\cite{dsendek_holistic_2017}, 
which could automatically detect combinations of descriptors and the intricate correlations between them,
although the lack of training data and interpretation of the results remain a major hurdle.

Consequently, the discovery of new ionic conductors has been driven up to now mostly by chemical intuition, as for example the discovery of the garnet family of structures~\cite{thangadurai_novel_2003} and of the argyrodites~\cite{deiseroth_li6ps5x_2008}, and incremental improvements of known ionic conductors, for example by anionic and cationic substitutions of known ionic conductors, as in the $\mathrm{Li_{4-x}Ge_{1-x}P_xS_4}$ thio-LISICON system~\cite{kanno_lithium_2001}, and equivalently in the $\mathrm{Li_{4\pm x}Si_{1-x}Y_xO_4 \quad (Y=Al,Ge,P)}$ LISICON system~\cite{deng_enhancing_2017}.

The varying and complex ionic diffusion mechanisms in diverse materials challenge the efforts to relate diffusion properties to simple descriptors, and this work explores a different approach. 
Instead of looking for descriptors, we try to directly compute
the bulk diffusion coefficients for every material, at a cost compatible with screening applications.
The goal is to combine the accurate framework of FPMD with physically motivated 
approximations that can tackle the time-limitations of this rather expensive technique.
This is achieved by simulating lithium ions in a potential energy landscape defined by the electrostatic and non-local interactions with a frozen host lattice and its charge density.
Li-ions moving through an environment of static obstacles
recall the game of pinball, and we refer to the model as the pinball model.
We will show that it correctly models the interaction of diffusing particles with the host lattice as well as the ion-ion interactions between diffusing particles, and is therefore a promising approach in the search for predictive models for ionic conductivity.
The pinball model does not account for lattice vibrations, and we discuss how much this limitation can influence the  results.
In addition, by comparing to first-principles simulations, we get novel insight on the correlation of lithium motion with the 
vibrations of the host lattice. 

In \autoref{sec-pinball} we present the model, while \autoref{sec-validation} and \ref{sec-results} discuss the validation strategy and results. The conclusions are presented in \autoref{sec-conclusion}.

\section{The pinball approximation}
\label{sec-pinball}

\begin{figure}[!t]
    \includegraphics[width=\hsize]{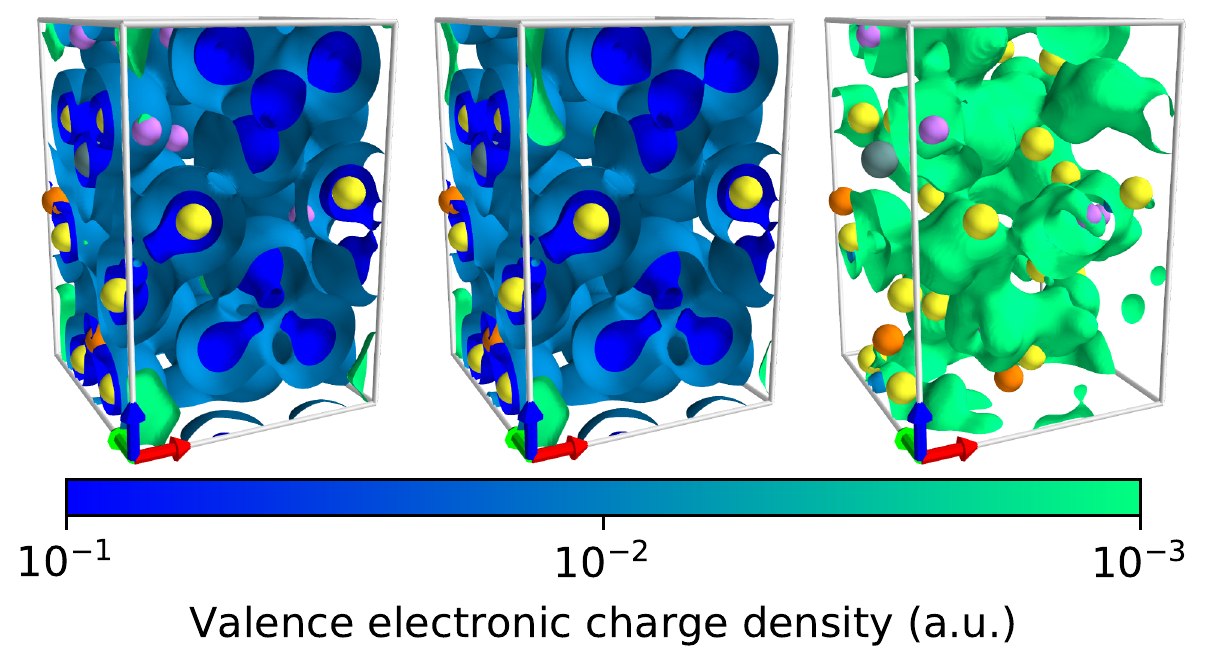}
       \caption{
           The left image shows a unit cell of $\mathrm{Li_{20}Ge_2P_4S_{24}}$ and 3 isosurfaces at $10^{-3}$, $10^{-2}$ and $10^{-1}$ of its ground-state valence electronic charge density.
           The center image displays the same isosurfaces for the charge density of  $\mathrm{Ge_2P_4S_{24}+20e^-}$,
           which is the former structure without the lithium cores but in the presence of lithium valence electrons.
           The same isosurfaces for the difference between the two previous charge densities are shown in the right image,
           showing that the error from the approximation is about two orders of magnitude lower than its characteristic values.
           Lithium, germanium, phosphorus and sulphur positions are shown as purple, olive, orange and yellow spheres, respectively, and the crystallographic directions a, b and c as red, green and blue arrows, respectively.
           }
       \label{fig.lgps-rho}
\end{figure}

We aim to model the diffusion of lithium ions through an ionic crystal. 
In these systems the ions
move through a host lattice containing anions of highly electronegative character,  such as oxygen, sulphur, nitrogen or halides.
Due to the large difference in electronegativity, the cations lose their 
2s-valence electrons, which are accommodated by the host-lattice anions, while keeping the 1s-states in their core configuration.
As a consequence, the valence electronic charge density depends weakly on the position of the diffusing cations. 
Such effect is shown in \autoref{fig.lgps-rho}, where the charge density of one of the fastest known Li-ionic conductors, $\mathrm{Li_{10}GeP_2S_{12}}$ (LGPS), is illustrated. 
We show the electronic charge density for one molecular dynamics snapshot computed in the explicit presence of lithium and compare it with the charge density obtained when all lithium ions are removed, while leaving their valence electrons in the simulation cell. 
The difference between the two charge densities is quite minor.
In order to exploit this behavior for modelling purposes it is convenient to separate 
the ionic conductor into two subsystems: 
\begin{itemize}
    \item A system of electropositive Li ions, 
          treated as electrostatic/quantum charges stripped of their valence electrons, but carrying a local and non-local effective pseudopotential
          taking into account the interaction of the entire electronic valence manifold with the Li 1s core states.
          All members of this system will be called \textit{pinballs}  $(P)$ in the remainder of this work.
    \item The host lattice $(H)$, consisting of the remaining, non-diffusing atoms and the valence electrons coming from the ionized Li; This is negatively charged.
\end{itemize}

We study the dynamics of the system using Born-Oppenheimer molecular dynamics (BOMD) 
in the framework of Kohn-Sham density functional theory (KS-DFT)~\cite{hohenberg_inhomogeneous_1964,kohn_self-consistent_1965}.
In the following, atomic units are used, where Planck's constant $\hbar$, the mass of the electron and the elemental charge are of unity.
For an ionic conductor with a sublattice $H$ of host ions at positions $R_H$ and a sublattice $P$ of pinballs 
at positions $R_P$, the Hamiltonian reads:
\begin{align}
\mathcal{H} = \frac{1}{2} \sum^H_h M_h \dot{R}_h^2 +  \frac{1}{2}  \sum^P_p M_p \dot{R}_p^2 + U (R_H, R_P) 
\label{eq.H1}
\end{align}
where the potential energy surface $U$ in the pseudopotential formulation of KS-DFT is given by:
\begin{align}
\notag U (R_H, R_P) =& E_N^{P-P} + E_N^{H-H} + E_N^{H-P} \\
\notag &+ \int n (\bm r) V_{LOC}^{H}(\bm r) d\bm r + \int n (\bm r) V_{LOC}^{P}(\bm r) d\bm r \\
\notag &+ \sum_i \braket{\psi_i | \hat{V}_{NL}^{H}| \psi_i } + \sum_i \braket{\psi_i | \hat{V}_{NL}^{P}| \psi_i } \\
 & + F[n]
\label{eq.U}
\end{align}
where $E_N^{A-B}$
is the electrostatic interaction between the nuclei dressed by the frozen core electrons of species $A$ and species $B$, $V^A_{LOC/NL}(r)$ are
the external local and non-local components of the pseudopotential of species $A$ that act on the valence electronic charge density $n(\bm r) = \sum_i \psi^*_i(\bm r)\psi_i(\bm r)$,
and where $F[n]$ is the universal functional of the charge density composed of the quantum kinetic energy operator,
the Hartree contribution and the exchange-correlation term:
\begin{align}
\notag F[n] =& - \frac{1}{2} 
\sum_i  \braket{\psi_i | \nabla^2 | \psi_i} + \frac{1}{2}\int  \int  \frac{n(\bm r) n(\bm r')}{|\bm r-\bm r'|} d\bm r' d\bm r\\
 &+ E_{XC}[n]
\end{align}
In Eq.~\eqref{eq.U}, we assume a negligible contribution from non-linear core corrections~\cite{louie_nonlinear_1982} and therefore do not account for the nonlinearity of the exchange and correlation interactions of the valence and core charge densities.

We apply two approximations to Eq.~\eqref{eq.U}, motivated by  physical intuition.
First, due to the weak dependence of the self-consistent valence electronic charge density on the Li-ion positions,
we approximate the fully self-consistent valence wave functions $\psi_i$ and valence charge density $n(\bm r)$ with the wave functions and charge density that are computed only in the presence of the host lattice, adding the additional electrons coming from the valence shell of the pinballs.
Technically, this results in a charged cell computation compensated by a neutralizing background.
We will denote the new wavefunctions and charge density with $\psi_{i, R_H}$ and $n_{R_H}$, respectively, because they depend parametrically solely on the host-lattice positions and 
are independent of the positions of the pinballs. 
Second, we neglect any motion of the host lattice and pin the host ions to their equilibrium positions $R_{H_0}$.
The application of these two approximations to Eq.~\eqref{eq.U} and the removal of all constant terms results in:
\begin{align}
    \notag \mathcal{H}_P =&  \frac{1}{2} \sum^P_p M_p \dot{R}_p^2 + E_N^{P-P} + E_N^{H-P}  \\
            &+ \int n_{R_{H_0}}(\bm r) V_{LOC}^P(\bm r)  d\bm r + \sum_i \braket{\psi_{i, R_{H_0}} | \hat{V}^{P}_{NL}| \psi_{i, R_{H_0}}}
    \label{eq.pinball-hamiltonian}
\end{align}
By definition $n_{R_{H_0}}$ and $\psi_{i, R_{H_0}}$ are time independent,
leading to a massive reduction of computational costs compared to FPMD, since the self-consistent 
calculations of $\psi_{i}(\bm r)$ and $n(\bm r)$ at every ionic step are eliminated and are calculated once, prior to the dynamics, in a single self-consistent calculation.

In order to improve further the accuracy of the model 
we introduce 4 phenomenological coefficients $\alpha_1,\alpha_2,\beta_1$ and $\beta_2$ in the Hamiltonian of Eq.~\eqref{eq.pinball-hamiltonian},
accounting for a potentially different screening of each contribution to the total energy due to charge polarization:
\begin{align}
    \notag \mathcal{H}_P =&  \frac{1}{2} \sum^P_p M_p \dot{R}_p^2 + \alpha_1 E_N^{P-P} + \alpha_2 E_N^{H-P} +\\
    \notag +&\beta_1 \int n_{R_{H_0}}(\bm r) V^{P}_{LOC}(\bm r)  d\bm r \\
           +&\beta_2 \sum_i \braket{\psi_{i, R_{H_0}} | \hat{V}^{P}_{NL}| \psi_{i, R_{H_0}}}
    \label{eq.pinball-screened}
\end{align}
The Hamiltonian framework is important for the resulting dynamics, permitting to extract 
dynamical properties under a well defined statistical ensemble. The corresponding forces are:
\begin{align}
    \notag \bm {\bm F_p}=&-\frac{d}{d\bm R_p} \left( \alpha_1 E_N^{P-P} + \alpha_2 E_N^{H-P}  \right) \\
    \notag &- \beta_1 \int n_{R_{H_0}}(\bm r) \frac{d}{d \bm R_p} V_{LOC}^P(\bm r)  d\bm r \\
    &- \beta_2 \sum_i \braket{ \psi_{i, R_{H_0}} | \frac{d \hat{V}_{NL}^P}{d \bm R_p}| \psi_{i, R_{H_0}}}
    \label{eq.forces1}
\end{align}
The coefficients are obtained from a force-matching procedure~\cite{ercolessi_interatomic_1994}; A standard multilinear-regression fit permits one to determine the 4 coefficients by minimizing the error with respect to exact KS-DFT forces in selected snapshots.
Deviation from unity of these parameters, as is generally observed, is due to the polarizability of the host matrix.
The fitting procedure is very inexpensive compared to the simulation times required for the computation of transport properties with FPMD, and technical details are discussed in App.~\ref{app-fitting}.

We note in passing that in the present work the last terms in Eq.~\eqref{eq.pinball-hamiltonian}, \eqref{eq.pinball-screened} or \eqref{eq.forces1} represents a norm-conserving pseudopotential, rather than an ultrasoft one or a PAW projector.
The extension to PAW and ultrasoft projectors, while feasible, is more cumbersome and provides a negligible advantage, 
since the cutoff for the charge density is broadly unaltered, and the additional efficiency in computing the ultrasoft or PAW projections due to a lower wave function cutoff is nullified by the larger prefactor in the calculation of the projection.

\section{Validation strategy}
\label{sec-validation}

We validated the pinball model in several systems characterized by different mechanisms of lithium diffusion and interactions
with the host lattice. For every system three statistical setups are simulated, associated to different approximations of the underlying dynamics, and compared against each other:
\begin{itemize}
\item
The ``free'' setup corresponds to standard FPMD of the full system, allowing both pinballs and the host lattice to move freely, with full self-consistency in the charge density.
\item 
In the ``constrained'' setup the host lattice is frozen in an equilibrium configuration, while the electronic charge density is allowed to relax self-consistently according to the instantaneous positions of the pinballs.
\item
Finally, in the ``pinball'' setup, with its Hamiltonian given by Eq.~\eqref{eq.pinball-screened}, 
any temporal variation of the electronic charge density and wave functions is neglected and replaced by $n_{R_{H_0}}$ and $\psi_{i, R_{H_0}}$.
\end{itemize}
For all three cases we compute diffusion properties under microcanonical evolution.
In addition to serve as a test bed, these simulations bring further physical insight into the diffusion mechanisms.
The comparison between the free and constrained dynamics allows for an assessment of the role of host lattice vibrations in lithium diffusion. 
Instead, comparing the constrained with the pinball setup enables us to quantify the importance of charge density fluctuations and self-consistency during the motion of lithium through the crystal.

We chose several systems from four different structural families to allow for general conclusions.
The first set of structures are represented by $\mathrm{Li_{10}GeP_2S_{12}}$ (LGPS) and derivatives, studied extensively with FPMD by Ong et al.~\cite{ong_phase_2012} and forming a set of highly conductive structures with variations in composition and volume.
As a second benchmark, we considered the LISICON structure $\mathrm{Li_{3.75}Si_{0.75}P_{0.25}O_4}$.
Unlike the  LGPS family, it shows a 3-dimensional conduction pathway~\cite{deng_structural_2015} 
for lithium, while still having a high conductivity that allows for treatment with full FPMD in reasonable timescales.
The third case is that of the layered vacancy conductor~\cite{huq_structural_2007,lapp_ionic_1983, alpen_ionic_1977} $\mathrm{Li_3N}$, very different from the LISICON and LGPS-like structures both in composition and morphology.
The high lithium content of $\mathrm{Li_3N}$ makes it an ideal testing case for the limits of the pinball model, since in this material 75\% of the atomic constituents are treated as pinballs.
In addition, the lower electronegativity of nitrogen (compared to oxygen or halides) suggests a lower degree of ionicity in this system when comparing to oxides, implying that lithium is more likely to affect its valence electron. 
Therefore $\mathrm{Li_3N}$ is included as a worst-case study.
Last, we included the non-conducting material $\mathrm{Li_3NbO_4}$,
since experiments by McLaren et al.~\cite{mclaren_li+_2004} show that undoped $\mathrm{Li_3NbO_4}$ is a poor ionic conductor, but also that the ionic conductivity increases upon doping with $\mathrm{Ni^{2+}}$.
A more detailed discussion on the selection of materials is given in App.~\ref{app-md-technicalities} together with technical details on the supercells used, chosen where necessary to reduce the effect of spurious correlations with periodic images.
For clarity, we will use the chemical formulas of the supercells in the remainder of this work.

The screening parameters needed to calculate energies and forces in the pinball model (Eq.~\eqref{eq.pinball-screened} and~\eqref{eq.forces1}) are obtained from a force-matching procedure.
For each material, several configurations from a training set are calculated fully self-consistently and in the pinball model without the screening parameters ($\alpha_1=\alpha_2=\beta_1=\beta_2=1$). Using least-squares linear regression we find the material-specific parameters that minimize the error of the pinball forces against self-consistent forces in this training set.
In this work, we take the training configurations to be the snapshots of the constrained simulation taken every $10ps$ of all the runs performed, since we did not observe any dependence of the parameters on the mean kinetic energy of the ensemble, i.e. the screening parameters are temperature independent.
We give additional details on the fitting procedure and the resulting parameters in App.~\ref{app-fitting}, together with an alternative procedure 
for the fitting that is of comparable accuracy but computationally much less expensive.

The ``free'' and ``constrained'' simulations were performed using the PWscf code of the Quantum ESPRESSO~\cite{giannozzi_quantum_2009} distribution. 
Dynamics in the pinball model required instead the development of an add-on functionality, in order to avoid the self-consistent cycle for the charge density optimization. 
For high-throughput capabilities, we wrote a plugin for the AiiDA materials' informatics platform~\cite{pizzi_aiida_2016} that is used in this work.
Technical details regarding the protocols for the molecular dynamics are given in App.~\ref{app-md-technicalities}, and
figures of merit regarding the computational speedup of the pinball model are presented in~App.~\ref{app-cpu-timings}.

\section{Numerical results and discussion}
\label{sec-results}

\begin{figure}[t]
\includegraphics[width=\hsize]{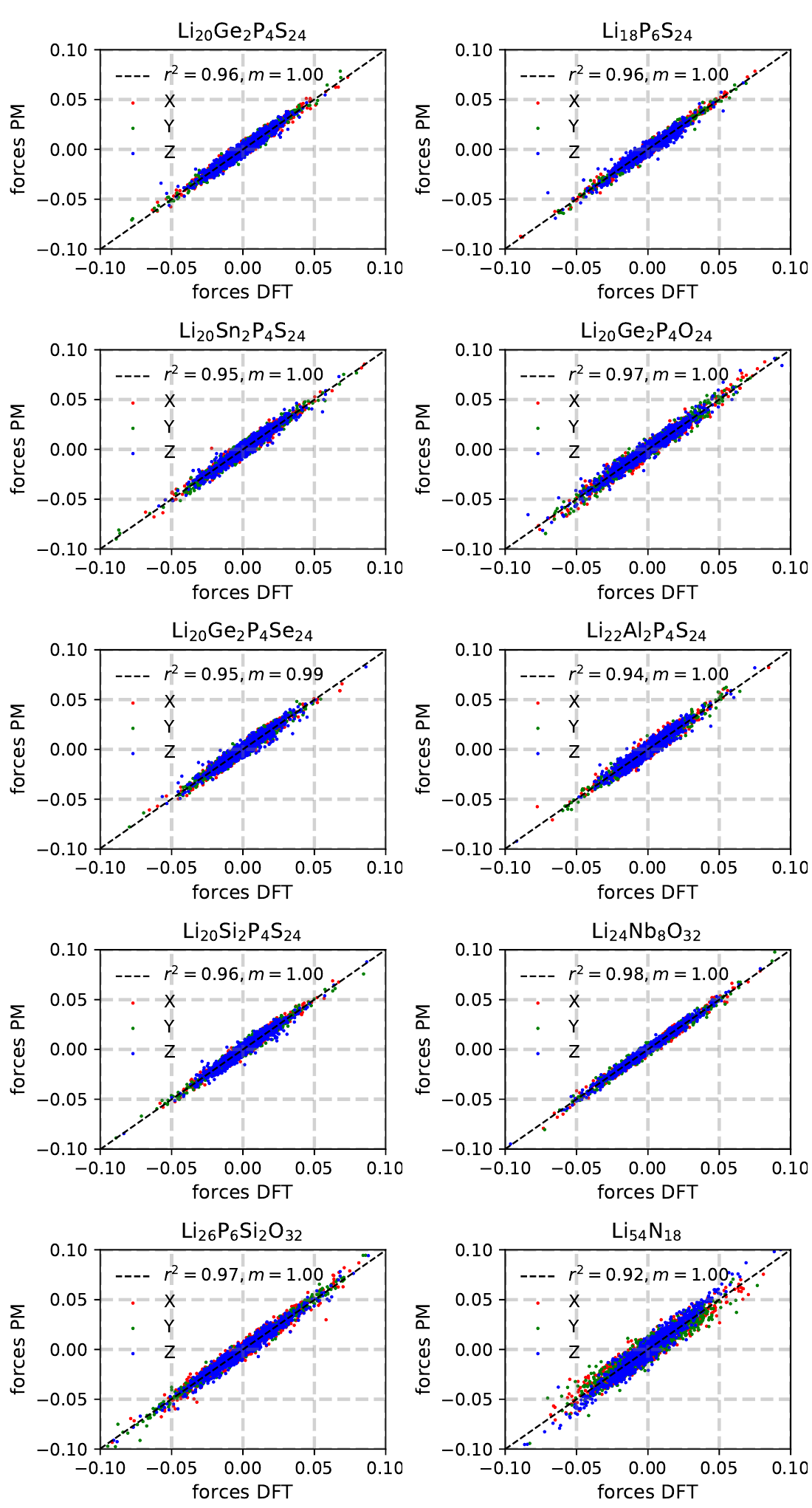}
\caption{
We show the forces (in Rydberg atomic units) in the pinball model on the y-axis against the forces calculated with KS-DFT on the x-axis.
Every point represent one component of the force vector for a lithium ion in a configuration from our validation set.
The best fit is shown as a dashed black line and in the legend its slope $m$ and the $r^2$ correlation coefficient are given as a quality measure of the fit.
}
\label{fig.forces}
\end{figure}

A good reproduction of Hellmann-Feynman forces, determining the time evolution of the system and \emph{a fortiori} the ensembles spanned, is a prerequisite for accurate dynamics.
In~\autoref{fig.forces} we show forces as resulting from the pinball model against those obtained with fully self-consistent calculations for all systems studied.
Configurations in this validation set originate from snapshots taken every time step from trajectories  calculated in the constrained setup at temperatures ranging from $600K$ to $1200K$.
Overall, forces in the pinball model are in excellent agreement with their DFT counterparts 
for all structures studied, indicating that the ``pinball'' setup can serve as a good 
approximation for the ``constrained'' one, for the temperature ranges spanned in this work.
The best fit ($r^2=0.984$) is produced for $\mathrm{Li_3NbO_4}$, as expected, since the lower polarizability of oxygen reduces the error made when freezing the electronic charge density.
The worst fit ($r^2=0.916$) can be seen in $\mathrm{Li_3N}$,
also expected because three quarters of the atoms in this structure are treated in an approximate way as pinballs.

Careful reproduction of the forces on the pinballs is a first step to show that the pinball model reproduces 
correctly static and dynamical properties.
In order to ensure that the model leads also to the correct distribution of the diffusing cation, 
we show the probability densities for the pinballs from each setup, calculated as:
\begin{equation}
 n_{\text{P}}(\bm r) =\left\langle \sum_{p}^{P} \delta(\bm r-\bm R_p) \right\rangle_{N,E,V}
\end{equation}
where the index $p$ runs over all pinballs in the simulation cell 
and angular brackets indicate a microcanonical mean over the molecular dynamics simulation.
In practice, the delta function is replaced by a gaussian of $0.2\AA$ standard deviation.

To estimate whether the vibrational properties of the pinballs are reproduced,
we calculate the vibrational density of states from the Fourier transform of the
velocity-velocity autocorrelation function $C_{\nu}(\omega)$:
\begin{equation}
C_{\nu}(\omega)= \frac{1}{N_P}\sum_p^{P} \int_{-\infty}^{+\infty}  \left\langle \bm V_p(t) \cdot \bm V_p(0) \right\rangle_{N,E,V} e^{i \omega t} dt
\end{equation}
where $\bm V_p$ is the velocity of a pinball $p$. 
In addition, the tracer diffusion coefficient is computed, which is a more delicate property to reproduce, strongly dependent on the time evolution of the system:
\begin{equation}
D_{tr}=\lim_{t \rightarrow \infty}  \frac{1}{N_P} \sum_p^{P} \frac{1}{6t}  \left\langle |\bm R_p(t)-\bm R_p(0)|^2 \right\rangle_{N,E,V}
\end{equation}
where $\bm R_p(t)$ is the position of a pinball $p$ at time $t$.
An estimate of the error of the tracer diffusion coefficients is obtained from a block analysis, with further details given in App.~\ref{app-md-technicalities}

A discussion of these figures-of-merit for the pinball model, ordered by material, follows.
The vibrational density of states and the isosurfaces shown in this work are always calculated from the simulation equilibrated,
as explained in App.~\ref{app-md-technicalities}, at $635K$, isovalues are reported in the respected caption.

\begin{figure}[t]
    \includegraphics[width=\hsize]{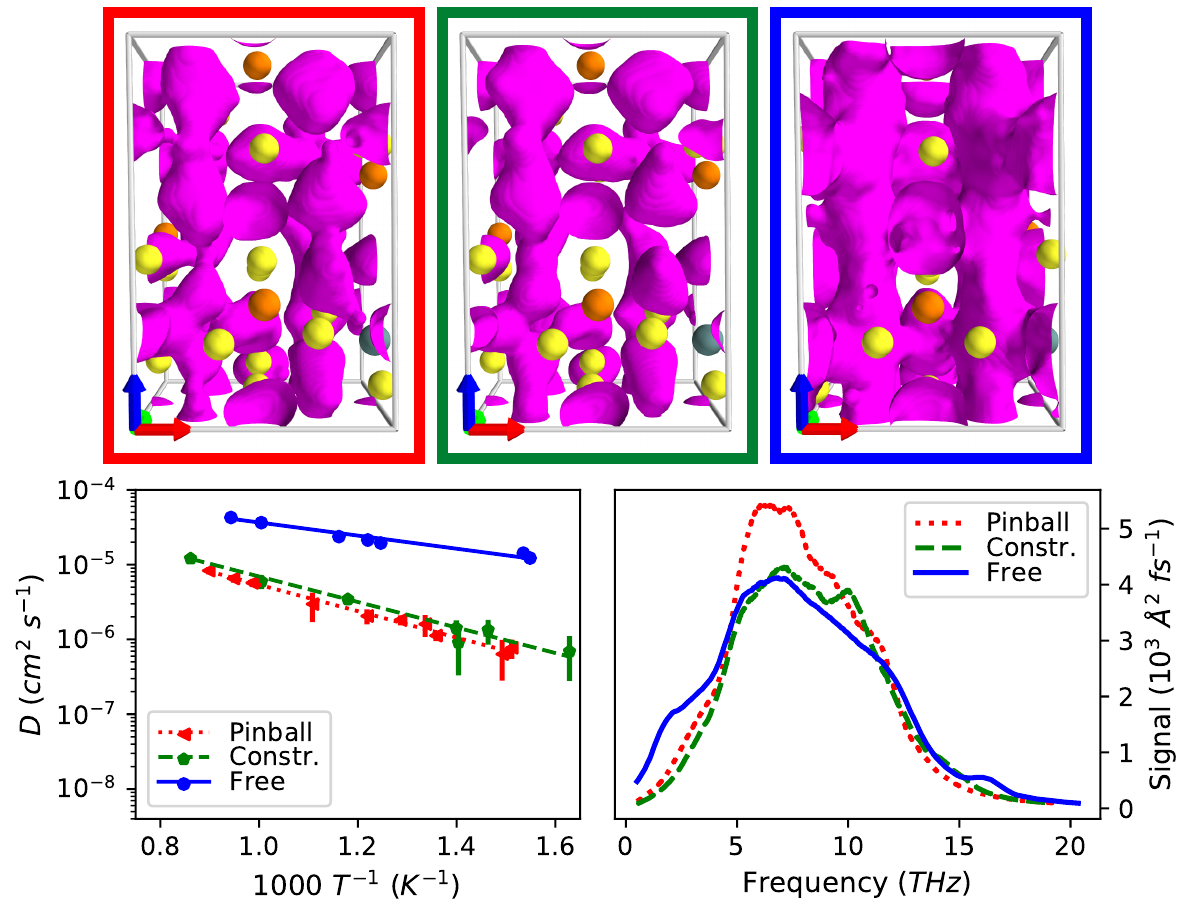}
\caption{
Results for $\mathrm{Li_{20}Ge_2P_4S_{24}}$: In the top row, the isosurfaces of the probability density of Li-ions for the same isovalue ($0.01\AA^{-3}$) are displayed:
on the left for the pinball model, in the center for the constrained setup and on the right for the free simulations.
Ge, P and S are represented as green, orange and yellow spheres, respectively, at their equilibrium position.
In the bottom left panel, the tracer diffusion coefficients are represented as a function of inverse temperature for the pinball, constrained and free setup with red dotted, blue dashed and green solid lines, respectively.
Error bars indicate the $2\sigma$-standard error of the mean.
On the bottom right panel we report the vibrational density of states of the Li-ions, with the same color coding.}
\label{fig.ionic-densities-Li20Ge2P4S24}
\end{figure}

\paragraph{The $LGPS$ family.}
\begin{figure}[t]
\includegraphics[width=\hsize]{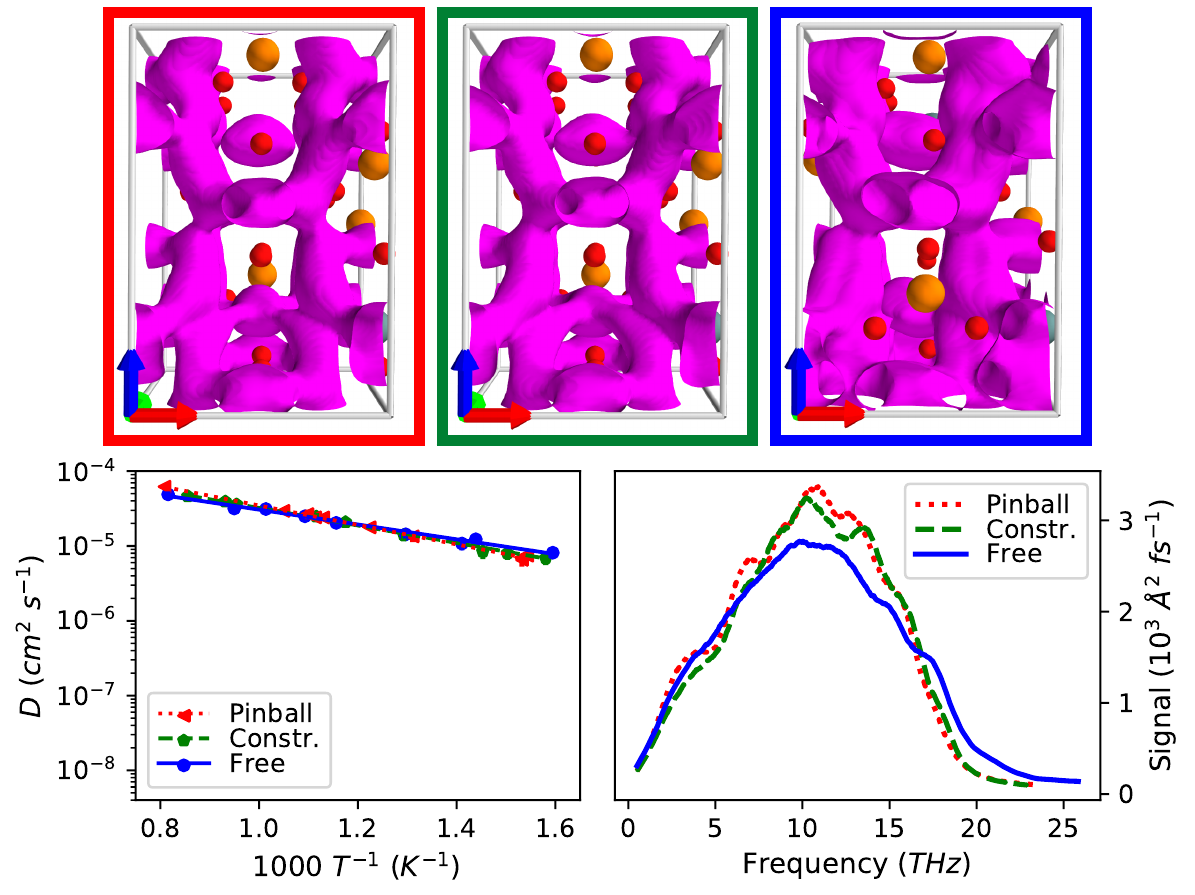}
\caption{
Results for  $\mathrm{Li_{20}Ge_2P_4O_{24}}$, as in \autoref{fig.ionic-densities-Li20Ge2P4S24},
with an isovalue of $0.01\AA^{-3}$ for the Li-ion probability density; Ge, P and O are represented as green, orange and red spheres, respectively, at their equilibrium position.}
\label{fig.ionic-densities-Li20Ge2P4O24}
\end{figure}

The tetragonal structure of $\mathrm{Li_{20}Ge_2P_4S_{24}}$, discovered by Kamaya et al.~\cite{kamaya_lithium_2011}, is 
well-known for its high and predominantly unidimensional transport. 
The isosurfaces of ionic density for the free setup, 
shown in the upper right panel in \autoref{fig.ionic-densities-Li20Ge2P4S24}, give clear evidence for the presence of unidimensional channels.
The same channels form in the constrained case, shown in the upper center panel, evidence that the topology of the carrier density is not affected by freezing the host lattice.
The ionic densities derived from the pinball, shown in the upper left panel, display very small differences when compared with the constrained setup, proving that the potential energy surfaces sampled in the two cases are very similar.
The bottom left panel displays the tracer diffusion coefficients calculated in the different setups and temperatures.
Lithium is more diffusive in the free setup than in the constrained one by about 
an order of magnitude in the temperature range studied, leading to an activation barrier of $0.33eV$ in the constrained setup against $0.17eV$ in the free setup.
Instead, the diffusion coefficients calculated in the pinball model agree well with the constrained simulations, with an activation barrier to diffusion of $0.35eV$.
In the bottom right panel we present the vibrational density of states for the Li-ions.
Apart from the $\omega \rightarrow 0$ limit, proportional to the diffusion coefficient,
the spectra show very good agreement, which becomes almost perfect when comparing the constrained and pinball setups. 
In summary, the pinball reproduces accurately dynamical and statical properties of the constrained setup in LGPS.
Freezing the charge density and switching to the pinball framework has a smaller effect on the resulting dynamics than constraining the movement of the host lattice.  We observed the same behavior for all sulphur and selenium derivatives.

$\mathrm{Li_{20}Ge_2P_4O_{24}}$ was obtained by Ong et al.~\cite{ong_phase_2012} by replacing sulphur in $\mathrm{Li_{20}Ge_2P_4S_{24}}$ with oxygen and relaxing the resulting cell.
We include this structure in the analysis due to the interesting differences with respect to LGPS.
Here, the isosurfaces in~\autoref{fig.ionic-densities-Li20Ge2P4O24} agree very well between the three different setups at each respective isovalue, as do
the diffusion coefficients and the vibrational density of states, without the differences between free and constrained setups observed in LGPS.

As discussed by Bachman et al.~\cite{bachman_inorganic_2016}, there is an understanding that the conductivity of a material can be enhanced by either softer vibrational modes or a higher polarizability of the host lattice that lithium is moving through.
The results on LGPS/LGPO suggest that the effect of freezing the host lattice has a significant effect for the sulphur containing compounds of the LGPS family, but not for their oxygen 
counterparts.

\begin{figure}[t]
\includegraphics[width=\hsize]{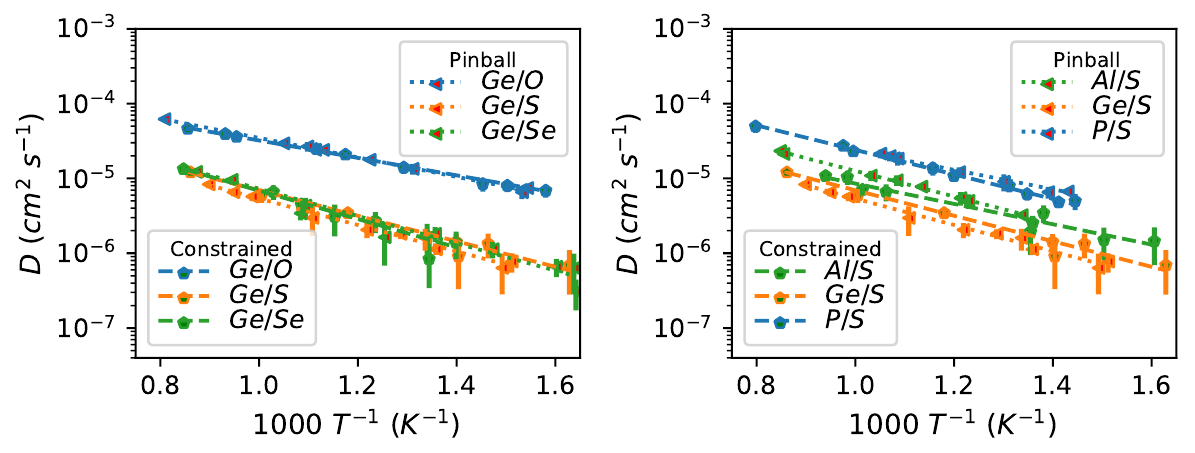}
\caption{Arrhenius behavior of the tracer diffusion coefficient for the constrained and the pinball setup (dashed and dotted lines of the same color). 
On the left for $\mathrm{Li_{20}Ge_2P_4S_{24}}$ and anionic substitutions with $\mathrm{O}$ and $\mathrm{Se}$, on the right for  $\mathrm{Li_{20}Ge_2P_4S_{24}}$ and aliovalent cationic substitutions.}
\label{fig-comparison-lgps}
\end{figure}

The LGPS-derivatives  studied by Ong et al.~\cite{ong_phase_2012} display small variations in the composition and volume, and the effect of these is discussed in the reference.
In order to estimate whether the pinball model correctly accounts for these variations, we plot the diffusion coefficients we obtained in this family in \autoref{fig-comparison-lgps}. 
The diffusion coefficients obtained in constrained setup span three orders of magnitude, and pinball simulations are able to reproduce the diffusion coefficients with remarkable accuracy
both for aliovalent and anionic substitutions, indicating that the model can account for subtle variations in lithium-ion density and anionic effects.
Display of the $\mathrm{Si}$ and $\mathrm{Sn}$-substitutions of $\mathrm{Ge}$ was omitted since no effect on the diffusion was found, regardless of the setup studied.

\FloatBarrier

\paragraph{$Li_{26}P_6Si_2O_{32}$.}
\begin{figure}[b]
\includegraphics[width=\hsize]{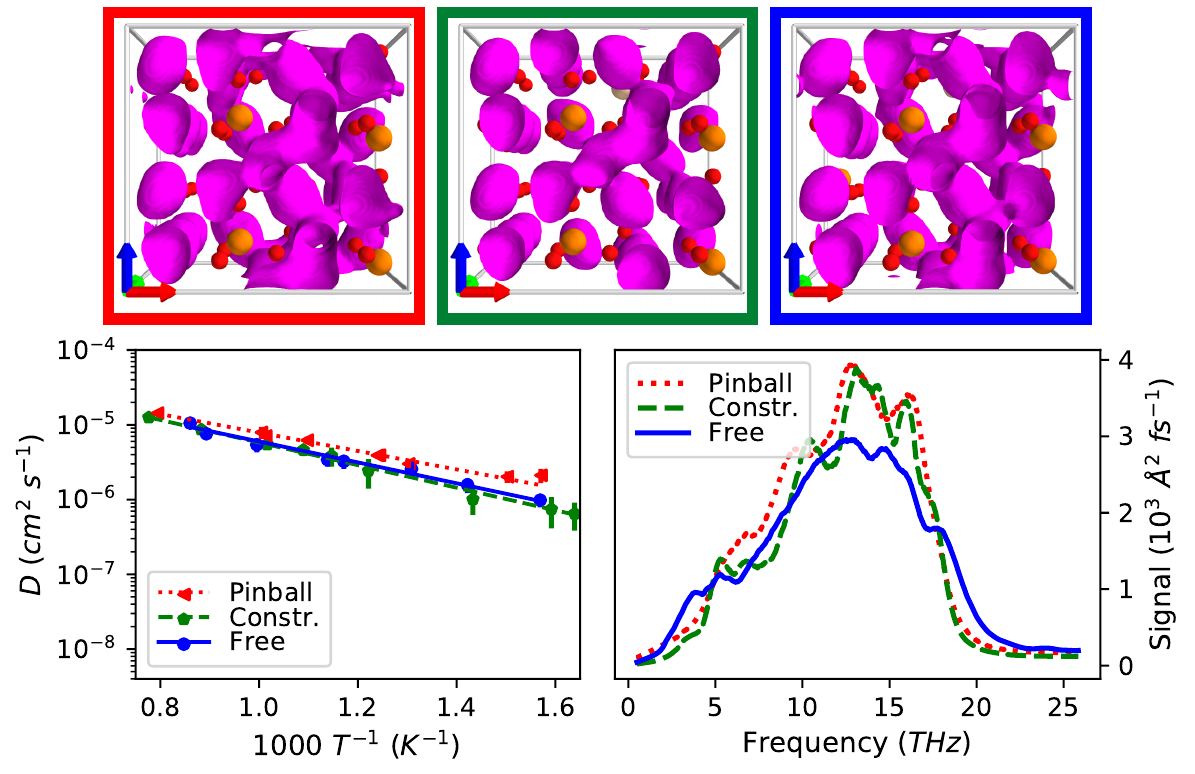}
\caption{
Results for  $\mathrm{Li_{26}P_6Si_2O_{32}}$, as in \autoref{fig.ionic-densities-Li20Ge2P4S24},
with an isovalue of $0.001\AA^{-3}$. Host lattice ions (P, Si, O) are shown at equilibrium as orange, yellow and red spheres, respectively.}
\label{fig.ionic-densities-Li26P6Si2O32}
\end{figure}

This LISICON compound was reported as a 3-dimensional conductor by Deng et al.~\cite{deng_structural_2015}, 
possessing the highest conductivities in the $Li_4SiO_4 - Li_3PO_4$ system.
Our FPMD results confirm that this material has a high conductivity and forms a 3-dimensional diffusion
network highlighted by the isosurfaces of the lithium probability density in~\autoref{fig.ionic-densities-Li26P6Si2O32}.
The pinball, constrained and free setups produce a very similar distribution of the lithium ions, as apparent from the shape of the isosurfaces.
The diffusion coefficients are in good agreement, although the pinball model produced marginally higher values.
Finally, the vibrational density of state in the pinball and constrained setups agree very closely:
a small discrepancy is observed between the free and the constrained setup, where some modes are softened.
Overall, similar to the case study of $Li_{20}Ge_2P_4O_{24}$, all setups are in very good agreement between each other.

\begin{figure}[t]
\includegraphics[width=\hsize]{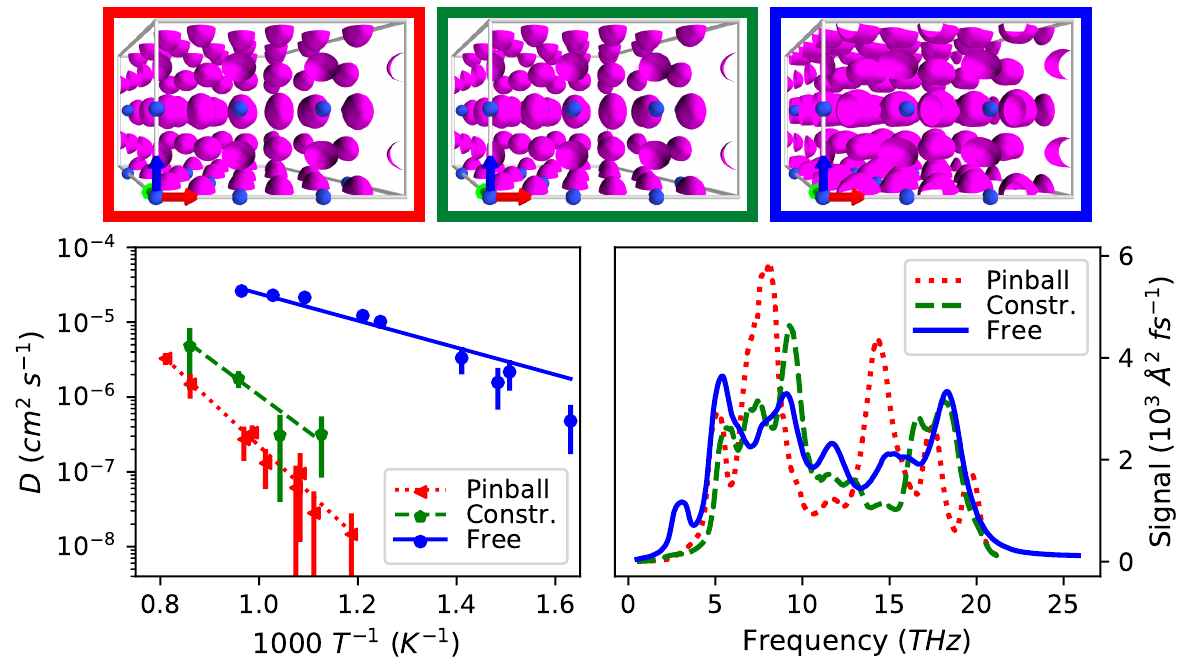}
\caption{
Results for $\mathrm{Li_{54}N_{18}}$, as in \autoref{fig.ionic-densities-Li20Ge2P4S24}. 
The isovalue for the Li-ion probability density is $0.05\AA^{-3}$, the host structure, consisting of nitrogen, is shown as blue spheres.
}
\label{fig.ionic-densities-Li54N18}
\end{figure}

\paragraph{$Li_{54}N_{18}$.}
This compound forms a layered structure of $\mathrm{Li_2N^-}$, with $\mathrm{Li^+}$ intercalated between the layers, resulting in 2-dimensional 
transport along mentioned layers~\cite{huq_structural_2007,lapp_ionic_1983, alpen_ionic_1977}.
The principal reason for inclusion of  $\mathrm{Li_{54}N_{18}}$ (a $3\times 3\times 2$  supercell of $\mathrm{Li_3N}$)
in this set is the high ratio of $Li$ to the respective anion $N$: 75\% of the constituents of this system are abstracted away in the pinball model.
In fact, in ~\autoref{fig.ionic-densities-Li54N18} discrepancies appear between the constrained and free setups in the ionic densities, 
diffusion coefficients and vibrational densities of states, highlighting how
motion of the lithium ions in this system is assisted by vibrations of the host lattice.
Nevertheless, the ionic densities in the constrained and pinball setup are in good agreement. 
Diffusion in the pinball model is underestimated when comparing with the constrained simulations, 
but the slopes of the logarithm of the diffusion with respect to inverse temperature are compatible. 
In summary, the pinball approximation reproduces reasonably well the constrained case, despite the fact that the charge 
density is obtained in the presence of just 25\% of the atomic constituents.
The pinball model is not able to reproduce the free setup due to the constraint of frozen anions, and not the constraint on the charge density.

\paragraph{$Li_{24}Nb_8O_{32}$.}

\begin{figure}[t]
\includegraphics[width=\hsize]{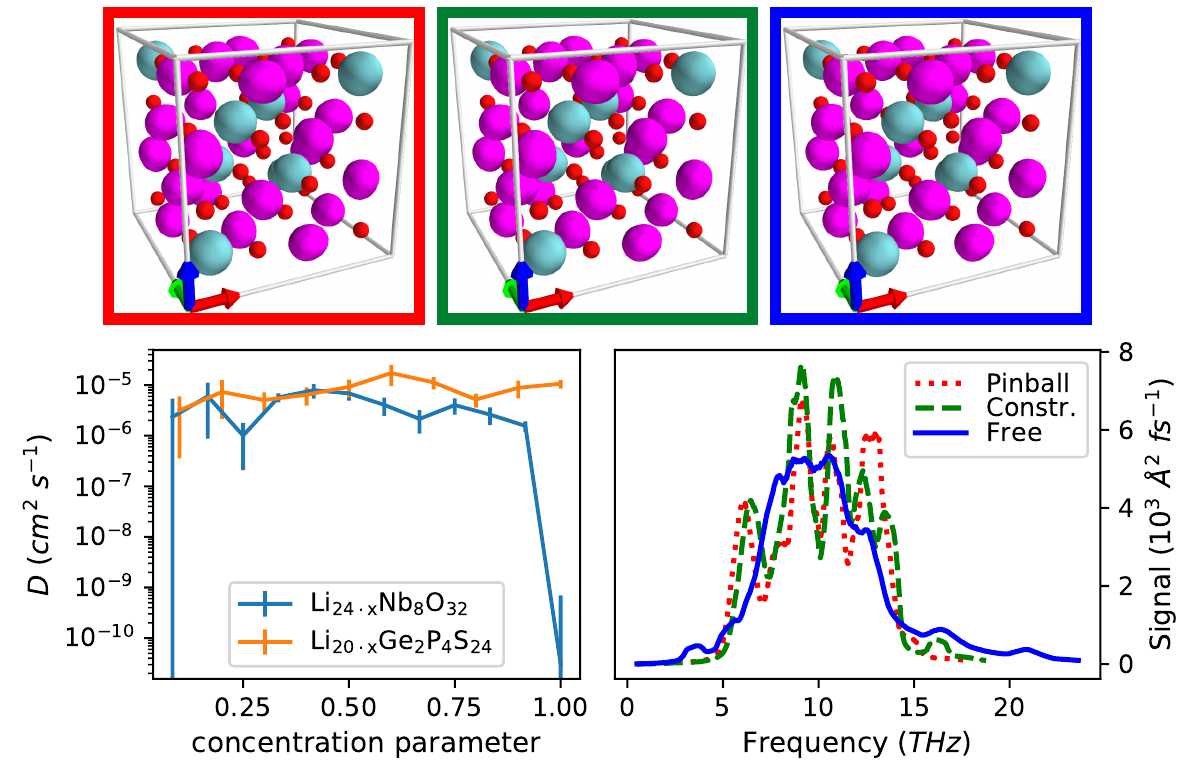}
\caption{
Results for $\mathrm{Li_{24}Nb_8O_{32}}$:  as in \autoref{fig.ionic-densities-Li20Ge2P4S24},
we plot in the top row the isosurfaces (at $0.05\AA^{-3}$) of the lithium-ion density for the pinball, constrained and free simulations, with the host structure of Nb and O illustrated as turquoise and red spheres, respectively, and the vibrational density of states in the bottom right panel.
In the bottom left panel we show the diffusion coefficient for $\mathrm{Li_{24\cdot x}Nb_8O_{32}}$ and $\mathrm{Li_{20\cdot x}Ge_2P_4S_{24}}$,
calculated in the pinball model at $1000K$, against the concentration parameter $x$ ranging from 0 (corresponding to no lithium) to 1 (fully lithiated structure).}
\label{fig.ionic-densities-Li24Nb8O32}
\end{figure}

This structure was refined by Grenier and Bassi~\cite{grenier_refinement_1965} and also by Ukei et al.~\cite{ukei_li3nbo4_1994} 
with a different space group but similar positions.
McLaren et al.~\cite{mclaren_li+_2004} reported the structure as a poor conductor, with
the conductivity sharply increasing  after doping with $Ni$, due to the creation of vacancies.
Our FPMD simulations of $\mathrm{Li_{24}Nb_8O_{32}}$ confirm that the structure is not conducting, as there is no diffusion of the lithium ions over 
the  observed time of roughly $200ps$, for all the simulations performed.
The  Li-ion densities, shown in \autoref{fig.ionic-densities-Li24Nb8O32}, agree perfectly between the pinball model, free and constrained setups and
suggest that the undoped compound is saturated in lithium, inhibiting vacancy-mediated ionic transport.
The vibrational density of states of the constrained setup and the pinball model are very similar, 
proving that the pinball model correctly captures the much more structured vibrational modes in this system.
As in previous materials, the distinct peaks of the vibrational spectrum in the constrained case are washed out 
when allowing the host lattice to move.

Instead of the Arrhenius behavior, which cannot be resolved due to the non-conducting nature of this material, the bottom left panel shows the estimated tracer diffusion coefficient
of $\mathrm{Li_{24\cdot x}Nb_8O_{32}}$ in the pinball model as a function of concentration.
The concentration of the pinballs can be changed without updating the charge density or the pinball parameters $\alpha_1$, $\alpha_2$, $\beta_1$ and $\beta_2$, if we assume that the only effect of the dopant atoms is to produce Li-ion vacancies to keep charge neutrality and that they do not affect the valence electronic charge density or the screening.
We vary the lithium concentration by removing lithium from the original structure, 
and calculate the diffusion coefficient of that partially delithiated structure at $1000K$.
As shown in the bottom left panel of \autoref{fig.ionic-densities-Li24Nb8O32}, the diffusivity increases sharply after the removal of lithium.
The same behavior is reported in McLaren's experimental study~\cite{mclaren_li+_2004}, where  doping with $Ni^{2+}$ sharply increases the ionic conductivity of this structure. 
Out of interest we repeated this exercise for LGPS, shown in orange, finding good diffusion at all concentrations.
The case of $\mathrm{Li_{24}Nb_8O_{32}}$ proves that the pinball model can correctly account for the effect of
variations in the concentration and compares qualitatively with experimental findings.
We expect the calculated diffusion coefficient in the niobate to be an overestimate
with respect to the experiments since the simulations do not capture the trapping
and channel blocking effect of the dopant, but the qualitative agreements suggests
that the pinball model can be used to study efficiently also the effects of concentration changes.

\paragraph{Interpretation}
The reported results can suggest general trends. 
For the three oxides studied in this work, the error made by the approximations of the model for the properties studied are quite minor. 
While cancellations of errors cannot be excluded, they are unlikely, since we compared different structures and different properties.
We conclude therefore that lithium-ion dynamics within a rigid solid-state structure of high ionic character are likely to be treated very accurately by the pinball model.
As one moves to less ionic systems, for example by replacing oxygen with sulphur, or moving to nitride systems, errors are introduced, 
as can be seen from a smaller $r^2$-value in the force comparison and a less accurate reproduction of the vibrational density of states and diffusion coefficients when comparing the pinball with the constrained setup.
However, this error is small compared to the error made by freezing the host lattice.
Already in LGPS, there is clear evidence for dynamic correlations between the anionic framework with the lithium ions.
An estimate of the ionic diffusion in the pinball model for structures that exhibit a close coupling between long-range diffusional modes and rotational 
modes of the host lattice, such as shown for \textit{closo}-borates~\cite{kweon_structural_2017}, is unlikely to work with the pinball model. 
The same can be concluded for the treatment of liquid systems, where the model can capture neither the correlations between anion and cation diffusion, 
nor the configurational degrees of freedom that can lead to enhanced diffusion.
Based on these considerations, we speculate that the frozen host approximation most often leads to an underestimate of transport coefficients, 
since more degrees of freedom give the system access to lower barriers, and because host-pinball dynamical correlations are neglected.
This variational behavior of the pinball model is compatible with its use as a screening criterion, where the estimate 
of the diffusion in the model can be seen as a lower bound for the actual diffusivity.

\section{Conclusions}
\label{sec-conclusion}

We proposed a ``pinball'' model to simulate efficiently the dynamics of lithium ions
at frozen host lattice and reported excellent agreement between this model and the
corresponding constrained FPMD simulations with regards to the topology of the carrier density,
characteristic vibrational frequencies and diffusion coefficients.
The qualitative behavior of the diffusion, as compared to fully unconstrained simulations,
is always reproduced, and non-diffusive materials can clearly be distinguished from diffusive ones; this makes the model suitable for screening applications.
An ongoing line of research is the extension of the model to allow for vibrations of the host lattice,
based on linear-response theory, leading to an even more faithful reproduction of the lithium-ion dynamics.

In addition, we show that the vibrations of the host lattice are an important contribution to the diffusion
coefficients of sulfide/nitride/selenide compounds, since constraining the host lattice leads to a decrease of conductivity by an order of magnitude in the 
temperature range studied for LGPS and sulphur/selenium derivatives, and several orders of magnitude for $\mathrm{Li_3N}$.
For the 3 oxides studied, we observe no significant effect from freezing the host lattice on the static or dynamical properties of lithium.
This observation suggests that the enhancement of lithium diffusivity in sulfides with respect to oxides is primarily due to 
different vibrational properties, especially the softer vibrational modes of the former, although this aspect could be investigated further.
Last, it is very unlikely that superionic conductivity in the compounds studied originates from complex bond rearrangements during the transitions,
since the pinball model cannot, by construction, account for fluctuations in the charge density, but is nevertheless able to predict accurately
the dynamical behavior of lithium ions in the frozen-lattice setup.

\section{Acknowledgement}
This research was supported by the Swiss National Science Foundation, through project 200021-159198 and the NCCR MARVEL.
We acknowledge computational support from the Swiss National Supercomputing Centre (CSCS).

\FloatBarrier

\bibliography{biblio}

\appendix

\section{Fitting procedure}
\label{app-fitting}
\begin{figure}[b]
\includegraphics[width=\hsize]{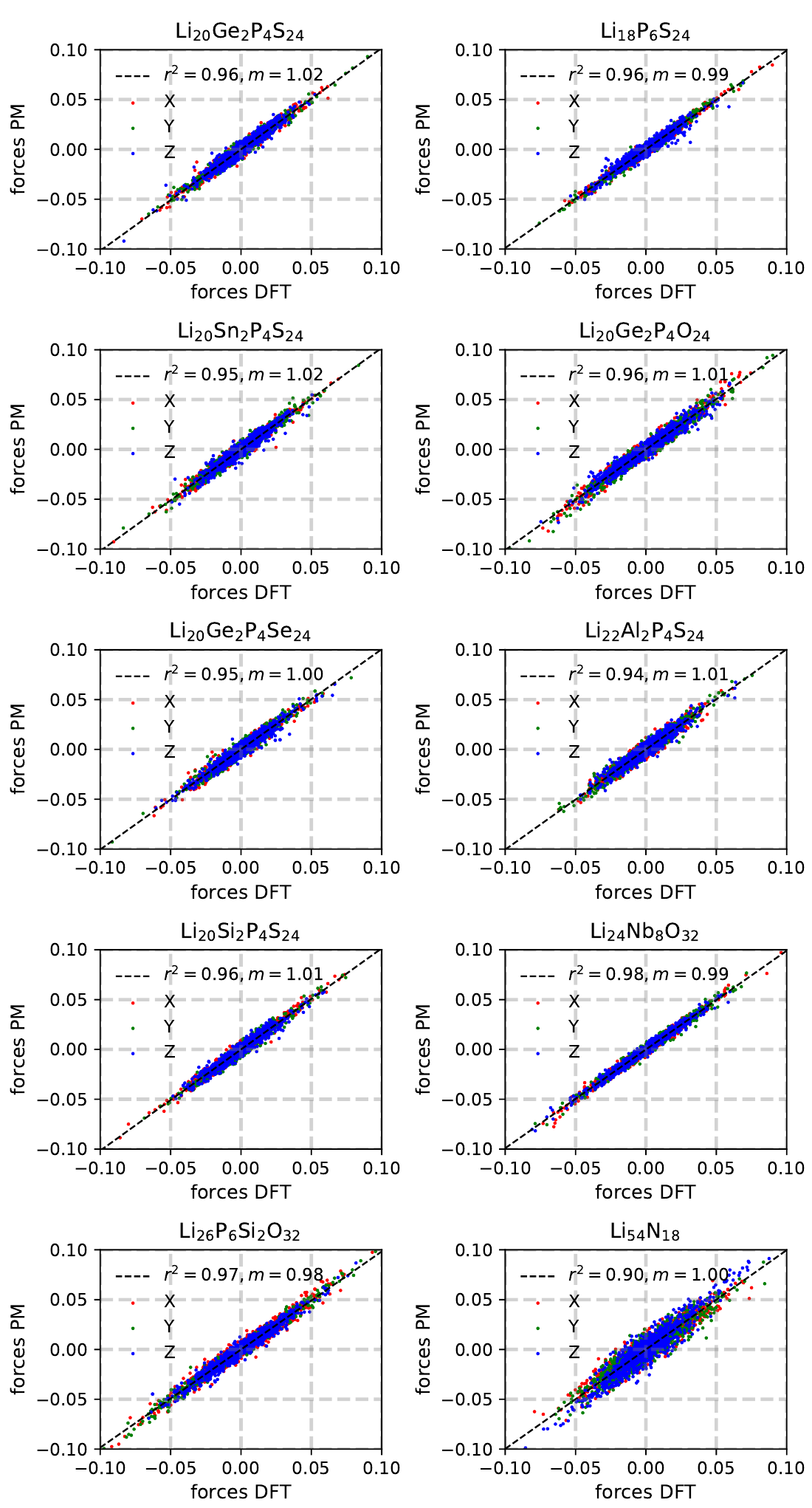}
\caption{
As in \autoref{fig.forces}, we show forces in Rydberg atomic units in the pinball model on the y-axis against the forces calculated with KS-DFT on the x-axis,
where the screening parameters have been determined using the smaller set discussed in App.~\ref{app-fitting}.
Quality of the fits remains excellent.}
\label{fig.cheap-forces}
\end{figure}

\begin{table}[b]
\begin{tabular}{lcccc}
Structure & $\alpha_1$ & $\alpha_2$ & $\beta_1$ & $\beta_2$  \\ \hline
$\mathrm{Li_{20}Ge_{2}P_{4}O_{24}}$ & 1.08481  & 2.18597  & 1.09664  & 0.52106 \\
$\mathrm{Li_{20}Ge_{2}P_{4}S_{24}}$ & 0.88117  & 1.50166  & 0.88354  & 0.32560 \\
$\mathrm{Li_{20}Ge_{2}P_{4}Se_{24}}$  & 0.85610  & 1.40126  & 0.85348  & 0.27956 \\
$\mathrm{Li_{20}Sn_{2}P_{4}S_{24}}$  & 0.95740  & 1.80620  & 0.96590  & 0.31865 \\
$\mathrm{Li_{20}Si_{2}P_{4}S_{24}}$ & 0.88896  & 1.55215  & 0.89317  & 0.33272 \\
$\mathrm{Li_{18}P_{6}S_{24}}$  & 0.90096  & 1.56215  & 0.90261  & 0.29648 \\
$\mathrm{Li_{22}Al_{2}P_{4}S_{24}}$  & 0.91360  & 1.62639  & 0.91756  & 0.34469 \\
$\mathrm{Li_{24}Nb_{8}O_{32}}$ & 1.14716  & 2.36787  & 1.16180  & 0.44543 \\
$\mathrm{Li_{26}P_{6}Si_{2}O_{32}}$  & 0.95224  & 1.76575  & 0.95931  & 0.52586 \\
$\mathrm{Li_{54}N_{18}}$ & 0.55120  & 0.31873  & 0.43687  & 0.58825 \\
\end{tabular}
\caption{Pinball parameters (used in this work).}
\end{table}

The parameters $\alpha_1$, $\alpha_2$, $\beta_1$ and $\beta_2$ in the pinball Hamiltonian \eqref{eq.pinball-hamiltonian} are 
determined by matching the forces for a training set of configurations between the pinball Hamiltonian and fully self-consistent calculations.
This is done by finding the parameters that minimize the error in a training set of size $M$:
\begin{align}
\notag S(\alpha_{1,2}, \beta_{1,2}) =&  \sum^M_k \sum_p^P \left\lVert F^{DFT}_{k,p} - \left[ -\frac{d}{d\bm R_{k,p}} \left( \alpha_1 E_N^{P-P} \right.\right.\right. \\
\notag	&+ \left. \alpha_2 E_N^{H-P}  \right) \\
    \notag &- \beta_1 \int n_{R_{H_0}}(\bm r) \frac{d}{d \bm R_{k,p}} V_{LOC}^P(\bm r)  d\bm r \\
     & \left.\left.   - \beta_2 \sum_i \braket{ \psi_{i, R_{H_0}} | \frac{d \hat{V}_{NL}^P}{d \bm R_{k,p}}| \psi_{i, R_{H_0}}} \right] \right\lVert^2
\end{align}
in a single shot using least-square regression.
The training set used in this work is a subset of the configurations from 
the constrained dynamics, selected every $10ps$ from our simulations.
We found no dependence of the converged parameters on the mean kinetic energy (i.e. temperature),
which enables us to use a large training set of configurations from all the simulations between $600K$ and $1200K$.

The primary reason to chose a large training set was to neglect any error coming from unconverged parameters, rather a need from the model itself.
For resource critical-applications it is not needed to run long dynamics to obtain uncorrelated snapshots to be used for fitting.
We investigated a faster fitting procedure:
Starting from the relaxed positions, we randomly displace the lithium-ions and create training configurations.
We observe the parameters of the model to be well converged when obtained from 100 random configurations of the respective supercells
with the pinball's displacements from equilibrium taken from a Gaussian distribution centered at $0$ and with a standard deviation of $0.1\AA$.
The forces we obtain for the validation set using this protocol are compared with DFT-forces in \autoref{fig.cheap-forces}, also showing very good agreement indicating that the 
this fitting procedure is equally good.
We also note that in resource-critical applications, the non-local projectors can be omitted, but the error made with this additional approximation needs to be assessed for each system.

\section{Molecular dynamics simulations}
\label{app-md-technicalities}
The structure of $\mathrm{Li_{10}GeP_2S_{12}}$ (LGPS) and its derivatives are taken from the Supplementary Information of Ong et al.~\cite{ong_phase_2012}.
The structure of $\mathrm{Li_{3.75}Si_{0.75}P_{0.25}O_4}$ as studied by Deng et al.~\cite{deng_structural_2015} comes from the corresponding ICSD entry 238600.
The structure of $\mathrm{Li_3N}$ is taken from COD entry 4311893 and that of $\mathrm{Li_3NbO_4}$ is taken from the Materials Project~\cite{jain_commentary_2013}, entry 31488.
Supercells are created by replicating the corresponding unit cells to ensure a minimum image distance of at least $6.4\AA$.
A subsequent relaxation of the host lattice geometry is performed in the absence of lithium ions.
The cell is not allowed to relax, so that the lattice vectors are still compatible with the reference structures.

Every system is thermalized at temperatures ranging from $600 K$ to $1200 K$, a standard choice in FPMD simulations of solid-state ionic conductors.
Thermalization is performed using a velocity rescaling thermostat~\cite{woodcock_isothermal_1971} for $20 ps$, 
after that it is switched off to recover microcanonical dynamics and to rule out any possible influence of the thermostat on the system dynamics.
Microcanonical simulation times vary according to the computational cost of each model.
In the case of the "free" and "constrained" first-principles simulations, $400-500 ps$ simulations are performed with a timestep of $1.93 fs$.
Half that timestep is used for the pinball model with simulation times of $750ps$.
We calculate the mean square displacements as a time average over all configurations in the microcanonical trajectories.
The diffusion coefficient is calculated from a linear fit of the mean square displacement between $2$ and $4ps$.
The error on the diffusion coefficient is estimated with a block analysis, where each trajectory is split into 8 independent blocks, each of at least $50 ps$.
The activation energies are estimated from a linear fit on the Arrhenius plot (logarithm of the  diffusion coefficient versus inverse temperature).

The exchange-correlation used in the DFT-calculations is PBE~\cite{perdew_generalized_1996};
pseudopotentials are those of the Standard Solid-state Pseudopotential (SSSP) library version 0.7 for efficiency,
with the recommended cutoffs~\cite{prandini_wwwmaterialscloudorg/sssp_2018},
with the exception of lithium for all pinball model simulations,
which required the construction of a custom pseudopotential using the atomic module of the Quantum ESPRESSO package.
This pseudopotential includes the 1s states in the core, and pseudizes the wave functions for the 2s and 2p states 
with a cutoff radius of $2.45 a.u.$ for both. Non-linear core corrections have not been included.

\section{CPU-timings}
\label{app-cpu-timings}

\begin{table}[b]
\begin{tabular}{lccc}
Structure & $t_{CPU} ^{BO} [s] $ & $ t_{CPU} ^{PM} [s] $ & $ t_{CPU}^{BO} /  t_{CPU} ^{PM}$ \\ \hline
$\mathrm{Li_{20}Ge_{2}P_{4}O_{24}}$ & $2.31 \cdot 10^{1}$ & $7.59 \cdot 10^{-2}$ & $3.05 \cdot 10^{2}$ \\
$\mathrm{Li_{20}Ge_{2}P_{4}S_{24}}$ & $3.18 \cdot 10^{1}$ & $1.07 \cdot 10^{-1}$ & $2.97 \cdot 10^{2}$ \\
$\mathrm{Li_{20}Ge_{2}P_{4}Se_{24}}$ & $3.26 \cdot 10^{1}$ & $1.30 \cdot 10^{-1}$ & $2.50 \cdot 10^{2}$ \\
$\mathrm{Li_{20}Sn_{2}P_{4}S_{24}}$ & $2.17 \cdot 10^{1}$ & $9.72 \cdot 10^{-2}$ & $2.23 \cdot 10^{2}$ \\
$\mathrm{Li_{20}Si_{2}P_{4}S_{24}}$ & $1.76 \cdot 10^{1}$ & $7.74 \cdot 10^{-2}$ & $2.28 \cdot 10^{2}$ \\
$\mathrm{Li_{18}P_{6}S_{24}}$ & $1.49 \cdot 10^{1}$ & $7.20 \cdot 10^{-2}$ & $2.07 \cdot 10^{2}$ \\
$\mathrm{Li_{22}Al_{2}P_{4}S_{24}}$ & $2.65 \cdot 10^{1}$ & $8.43 \cdot 10^{-2}$ & $3.14 \cdot 10^{2}$ \\
$\mathrm{Li_{24}Nb_{8}O_{32}}$ & $3.07 \cdot 10^{1}$ & $1.18 \cdot 10^{-1}$ & $2.59 \cdot 10^{2}$ \\
$\mathrm{Li_{26}P_{6}Si_{2}O_{32}}$ & $1.86 \cdot 10^{1}$ & $1.19 \cdot 10^{-1}$ & $1.56 \cdot 10^{2}$ \\
$\mathrm{Li_{54}N_{18}}$ & $3.21 \cdot 10^{1}$ & $2.85 \cdot 10^{-1}$ & $1.13 \cdot 10^{2}$ \\
\end{tabular}
\caption{CPU time $ t$ per ionic step is given for Born-Oppenheimer MD and for the pinball model dynamics in columns 2 and 3, respectively, for each structure.
The last column shows the ratio of the timings and represents the computational speedup of the pinball model with respect to DFT-based BOMD. The timings are normalized by the number of nodes used in the respective calculation, assuming linear scaling.}
\label{tab-cpu-timings}
\end{table}

To apply pinball model in a screening scenario, the computational cost of simulating time evolution via molecular dynamics is of paramount importance.
In~\autoref{tab-cpu-timings} we report the average CPU time per node for a single ionic step in the pinball model and compare this to BOMD 
for all calculations that have been used in this work.
The speedup of the pinball model is at least 2 orders of magnitude for every structure studied.
All calculations are run on an Intel Xeon cluster, on 1 or 2 nodes each, 
equipped with 2 Ivy Bridge processors with 8 cores each.
The computational efficiency is mostly due to avoiding a call to the routines for the self-consistent minimization.
Further improvements stem from avoiding the recalculation of structure factors for non-diffusive species,
and the recalculation of the reciprocal-space charge density before evaluating the forces and total energy.
Whereas the calculation of the forces is parallelized with MPI, the propagation of ions 
is still done in serial. 
Together with other possible optimizations, the efficiency of the pinball model can most likely be further improved, 
and ~\autoref{tab-cpu-timings} should be seen as a lower bound for the speedup of the model.
Regarding scaling, the pinball model scales linearly with the number of local and
quadratically with the number of non-local projectors,
and linearly with the grid size for the charge density and wave functions.
Therefore, a worst-case cubic scaling with system size at constant grid point density and Li-stochiometry is found with non-local projectors,
and quadratic scaling if only local projections are included.
\end{document}